\documentclass[prb,twocolumn,showpacs,superscriptaddress,amsmath,amssymb]{revtex4}

\usepackage{graphicx}% Include figure files
\usepackage{amssymb,amsmath}
\usepackage{dcolumn}% Align table columns on decimal point
\usepackage{bm}% bold math
\usepackage{mathrsfs}% bold math
\usepackage{color}

\begin{document}

\title{Electric generation of vortices in an exciton-polariton superfluid}

\author{H. Flayac}
\affiliation{Institut Pascal, PHOTON-N2, Clermont Universit\'{e}, Blaise Pascal University, CNRS, 24 avenue des Landais, 63177 Aubi\`{e}re Cedex, France}
\author{G. Pavlovic}
\affiliation{International Institute of Physics, Av. Odilon Gomes de Lima, 1722, CEP 59078-400 Capim Macio Natal, RN, Brazil}
\author{M. A. Kaliteevski}
\affiliation{Academic University - Nanotechnology Research and Education Centre, Khlopina 8/3, 194021, St. Petersburg, Russia}
\affiliation{Ioffe Physical-Technical Institute, Polytekhnicheskaya 26, 194021, St. Petersburg, Russia}
\author{I. A. Shelykh}
\affiliation{Physics Department, University of Iceland, Dunhagi-3, IS-107, Reykjavik, Iceland}

%\date{\today}% It is always \today, today,
             %  but any date may be explicitly specified

\begin{abstract}
We have theoretically demonstrated the on demand electric generation of vortices in an exciton-polariton superfluid. Electric pulses applied to a horseshoe-shaped metallic mesa, deposited on top of the microcavity, generate a non-cylindrically symmetric solitonic wave in the system. Breakdown of its wavefront at focal points leads to the formation of vortex-antivortex pairs which subsequently propagate in the superfluid. The trajectory of these vortex dipoles can be controlled by applying a voltage to additional electrodes. They can be confined within channels formed by metallic stripes and unbound by a wedged mesa giving birth to grey solitons. Finally single static vortices can be generated using a single metallic plate configuration.
\end{abstract}
 \maketitle

\section{Introduction}
Elementary excitations in a superfluid Bose-Einstein condensates\cite{GPBEC} (BEC) are quantum vortices, topological defects arising from the Onsager-Feynman quantization rule for the flow\cite{OnsagerFeynman}:
\begin{equation}
m^\ast\oint{\vec{v}_s(\vec{r},t)d\vec{r}}=l h, \label{eq1}
\end{equation}
where $h$ is the Plank constant, $m^{*}$ is the effective mass of the particles involved and $\vec{v}_s(\vec{r},t)=\hbar/m^{*}\vec{\nabla}{\phi(\vec{r},t)}$ is the velocity field of the collective state characterized by the order parameter
$\psi(\vec{r},t)=n(\vec{r},t)e^{i\phi(\vec{r},t)}$. The winding number $l$ is an integer that reflects the quantization of the angular momentum of the superfluid flow. A single vortex is characterized by a vanishing density and therefore by a singularity of the phase $\phi$ in its center. The spatial dimension of a vortex is defined by the healing length $\xi^2=\hbar^2/\alpha n m^\ast$ of the condensate, where $n$ is the concentration of the superfluid and $\alpha$ is the self-interaction constant.

Experimental investigations of the formation of vortices remain one of the topical issues in the cold atoms physics. In the first related experiment\cite{Matthews} a kind of phase-imprinting technique, originally proposed in Ref.\onlinecite{Williams}, was implemented in order to generate a velocity field satisfying the condition given by Eq.(\ref{eq1}). Besides, the system as a whole can be put into rotation in order to transfer angular momentum to individual vortices: the so-called rotating bucket method. Arrays of vortices appear in the ground state provided that the angular velocity of the fluid surpasses a critical value. Despite being a well-established technique regarding superfluid Helium\cite{Yarmchuk}, it showed poor efficiency for cold atomic BECs\cite{revKasamatsu}. Another famous strategy for the creation of vortices and vortex lattices consists in the stirring of a trapped BEC of $^{87}$Rb atoms with a laser beam\cite{Madison}.

The onset of BEC and superfluidity is not restricted to atomic systems. Indeed, these quantum coherent phenomena were predicted or reported for a variety of condensed matter systems, including indirect excitons\cite{IndirectEx}, magnons\cite{Magnons} and their spin\cite{MagnonsSSF} and exciton-polaritons\cite{Kasprzak,Balili,Amo} (polaritons). The latter system is especially promising. Exciton-polaritons are 2D quasi-particles massively generated when quantum well excitons and cavity photons become strongly coupled inside a planar microcavity\cite{KavokinBook}. Due to their composite nature, polaritons possess a number of properties that distinguish them from other quasiparticles in condensed matter systems. First, in the low density limit they behave as an interacting Bose gas. Second, the presence of a photonic component in their wavefunction makes their effective mass extremely small ($10^{-4}-10^{-5}$ the electron mass), while the excitonic component makes polariton-polariton interactions possible. Consequently, the polaritonic system can undergo a thermalization and a transition to a state similar to BEC even at room temperature\cite{Christopoulos,Levrat}. As polaritons are 2D objects, this transition has an universality class of Berezinskii-Kosterlitz-Thouless transition\cite{PitaevskiiStringari} and is accompanied with the formation of vortex-antivortex pairs (VAPs) (or vortex dipoles) in the vicinity of some critical temperature\cite{Hadzibabic}. The rapid development of experimental studies of polaritonic systems has allowed to make a major step forward in the study of vortices. Their spontaneous formation pinned to defects was reported under a non-resonant pumping configuration\cite{KLagoudakis}. As well, the nucleation of vortices was observed due to the inhomogeneity of the pump laser spot\cite{Roumpos}. In addition, an artificial generation of vortices was achieved under resonant pumping in the regime of polariton optical parametric oscillator\cite{Marchetti,Tosi} using a Gauss-Laguerre probe to imprint the proper singular phase, and in experiments involving a polariton condensate flowing against defects, intrinsic or extrinsic to the microcavity\cite{NardinV,SanvittoV,GrossoV}.

In this paper we propose a novel strategy for the generation of vortices in the polariton superfluid. Contrary to previous proposals, the vortices are not excited optically but electrically. The idea is based on the possibility of creating dynamical confining potentials for the particles, by deposition of metallic contacts on top of the sample and by the application of short voltage impulses to them. We will show that the use of a horseshoe-shaped contact [shown in Fig.\ref{Fig1}] leads to the generation and propagation of VAPs. Such a structure acts as a "vortex gun" triggered by an applied external potential. On the basis of this main element, we will propose schemes towards the guiding and unbinding of VAPS provided by additional metallic gates. In addition, we will discuss another simple configuration allowing the nucleation of single vortices.

\section{The Model}
\begin{figure}[tbp]
\includegraphics[width=0.99\linewidth]{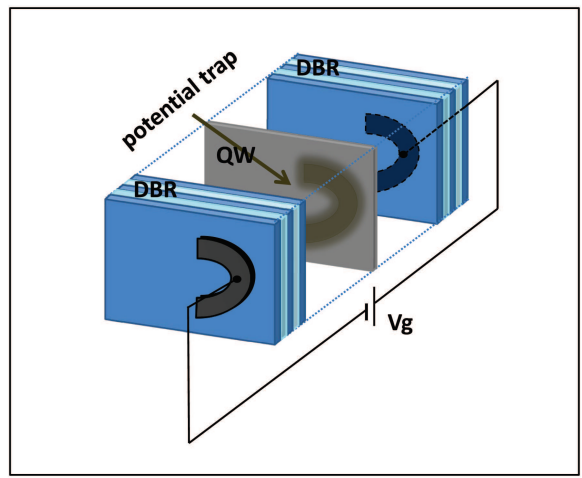}
\caption{(Color online) Vortex gun: A microcavity with a horseshoe-shaped metallic contact deposited on top of its DBRs. The metal creates a local potential trap for polaritons. The application of a time-dependent voltage $V_g$ is able to increase temporally the depth of the trap towards the generation of VAPs.} \label{Fig1}
\end{figure}

We will consider a microcavity, assumed to have a high quality factor and therefore, characterized by weak intrinsic disorder (e.g. GaAs based), with metallic mesas deposited on its top. A voltage can be applied to the metal via integrated electric contacts [see Fig.\ref{Fig1}]. We assume that the polariton condensate is non-resonantly populated by means of a far detuned \emph{cw}-homogenous pump over the whole sample. This is a crucial point in so far as, a resonant injection scheme would continually impose, at any time, a well defined phase to the condensate preventing phase singularities (vortices) from appearing. The polariton population $n$ can remain constant and fix a chemical potential for the system, provided that a quasi-equilibrium regime between gains -- from excitons that relax towards the ground state -- and losses -- from photons escaping through the Bragg mirrors -- is achieved.

The role of the metallic contacts on top of the sample is twofold: First, at the interface between a metal and dielectric Bragg reflectors (DBRs), Tamm plasmon-polaritons states can appear. Such events lead to a local \textit{redshift} of the lower polariton modes and thus a time-independent potential \textit{trap} is created in the region where the metal is deposited\cite{KaliteevskiPRB,Symonds}. We note that this configuration is different from the one involving surface states of a Shottky-like junction resulting in a \textit{blueshift} of the polariton modes\cite{Yamamoto,FlayacBOs}. The depth of the trap depends on the type of metal and on the thickness of the layer and values of several meVs are easily achievable. Second, the application of a voltage to the metallic contacts gives birth to an additional \emph{redshift} of the polariton energy due to the excitonic Stark effect\cite{Kaliteevski,Liew}. The value of the shift depends of course on the applied voltage and values in the range of 0-50 kV result in a potential $U_V$ lying between -2 and 0 meV. Besides, the abrupt appearance of a potential well, can strongly perturb the condensate locally in the region of the contacts. Such a perturbation results in the excitation of a nonlinear density wave similar to a ring soliton\cite{RingSolitons} and possibly of dispersive shock waves in the condensate\cite{Hoefer}. Vortices can emerge, when the solitonic wave expand above some critical velocity producing a vortex necklace\cite{RingSolitons} or when two nonlinear waves collide and interfere. This latter situation can be realized using two facing electric contacts or just a single one with the proper geometry as we will see.

For the description of the dynamics of the system, we have used the standard Gross-Pitaevskii equation\cite{GPBEC} for the macroscopic wavefunction of the condensate $\psi(\vec{r},t)$:
\begin{eqnarray}
\nonumber i\hbar \frac{{\partial \psi (\vec r,t)}}{{\partial t}} =  &-& \frac{{{\hbar ^2}}}{{2{m^*}}}\Delta \psi (\vec r,t) + \alpha
|\psi (\vec r,t){|^2}\psi (\vec r,t) \\
  &+& \left[ {{U_0}(\vec r) + {U_V}(\vec r,t)} \right]\psi (\vec r,t)
\end{eqnarray}
$U_0(\vec{r})$ is a potential well of the amplitude of $2$ meV created by the metallic contacts and appearing due to the formation of Tamm plasmon-polaritons. $U_V(\vec{r},t)$ is an extra time dependent potential accounting for the exciton Stark shift generated by the applied voltages. $m^\ast=5\times10^{-5}m_0$ ($m_0$ is the electron mass) is the polariton effective mass. $\alpha=6 x E_b a_B^2$ -- where $x=0.5$ is the excitonic fraction of polaritons, $E_b=10$ meV is the exciton binding energy and $a_B=10^{-2}\mu$m its Bohr radius -- is the polariton-polariton interaction constant. The nonresonantly populated driven-dissipative polariton condensate could be more accurately described by a model including finite lifetime of the particles and coupling of the condensate with the non-resonantly pumped excitonic reservoir\cite{WCModel}. We have therefore checked [see Appendix] with the latter model that the VAP generation described below remains qualitatively the same as the one obtained with the simpler Eq.(2), which gives full legitimacy to our simplified model. We have also neglected the spin degree of freedom of polaritons assuming that the polariton condensate is linearly polarized\cite{Balili} and that both spin components are fully equivalent. One should keep in mind, however, that the account of the spin of polaritons can in principle have important consequences for their dynamics\cite{ShelykhReview} as more complex types of topological excitations, such as half-vortices\cite{Rubo,Flayac,Lagoudakis} or half-solitons\cite{FlayacHalfSol} could be excited in the spinor system. Consideration of these interesting effects lies beyond the scopes of the present paper and will be a matter of future works.

\section{The vortex gun}
\begin{figure}[tbp]
\includegraphics[width=0.99\linewidth]{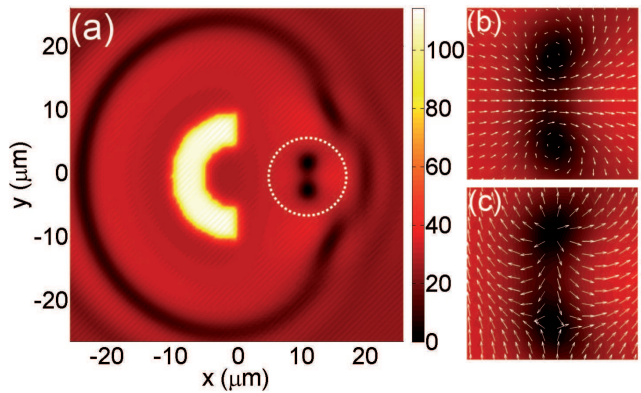}
\caption{(Color online) VAP nucleated by the application of a voltage impulse of amplitude $-1.5$ meV and duration 5 ps to the horseshoe contact. (a) Density $n(x,y)$ ($\mu$m$^{-2}$) of the condensate, 20 ps after the pulse. One clearly sees the solitonic density wave propagating away from the contact (dark regions) and the single VAP (dashed white circle). (b) Velocity field of the condensate and (c) phase vector field $(\cos \phi, \sin \phi)$.} \label{Fig2}
\end{figure}

We start our description with the configuration shown at Fig.\ref{Fig1}. A metallic contact in the form of a horseshoe (half a ring) is deposited on top of DBRs. It has an inner radius of $5$ $\mu$m (roughly 2$\xi$) and outer radius of $10$ $\mu$m. We apply to the contacts a short 5 ps long Gaussian voltage impulse, producing a time-dependent redshift of the amplitude of $1.5$ meV associated with the potential well $U_V$. It results in a local non-adiabatic perturbation of the condensate which generates a non-cylindrically symmetric solitonic density wave propagating outwards from the contact. In the inner part of the horseshoe the breakdown of the meeting wavefronts occurs, allowing the nucleation a VAP propagating along the axis of symmetry of the system, as shown in Fig.\ref{Fig2}. The vortices are evidenced by: the vanishing density at their core visible in the colormap, the characteristic velocity field tangent to them in the panel (b) and the winding of the phase in anticlockwise (vortex $l=+1$) and clockwise (antivortex $l=-1$) direction around their core in the panel (c). The distance separating the vortex and antivortex in the pair related to their speed of propagation depends on the size of the contact, on the applied voltage and on the concentration of the polariton superfluid (here on average $\overline n=2.5\times10^{9}$cm$^{-2}$).
We have therefore proposed a system able to generate VAPs on demand propagating along straight lines: a "vortex gun". Moreover, the modification of the excitation condition allows the creation of more than a single VAP. For example if a periodic sequence of voltage impulses is applied to the horseshoe, the generation of trains of VAPs is observed [see Ref.\onlinecite{Movie1} for a movie].

\section{Vortex guiding}
\begin{figure}[tbp]
\includegraphics[width=0.99\linewidth]{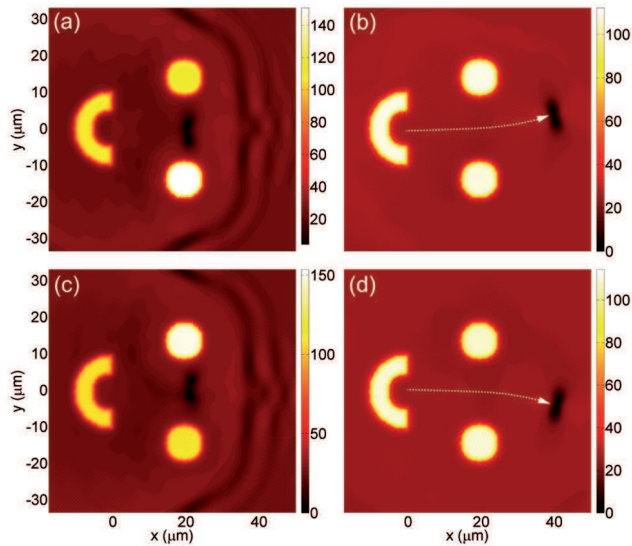}
\caption{(Color online) Control of the VAP trajectory by a pair of circular gates. Panels (a)-(b) and (c)-(d) show the density of the polariton superfluid in real space 30 ps and 55 ps after the VAP generation respectively. The upper (lower) panels correspond to a voltage applied to the lower (upper) mesa and thus to an upward (downward) deviation of the VAP. The corresponding trajectories are marked by the white dashed arrows.} \label{Fig3}
\end{figure}

The motion of vortices is affected by flow directions and density gradients\cite{Aioi}. Indeed, as the metal creates a potential well for polaritons, it makes the concentration of the condensate under the contacts greater than in the surrounding area [it is clearly visible e.g. in Fig.\ref{Fig2}(a)]. We note that, even if photoabsorption from the nonresonant pump laser can be blocked by the presence of the metal, the polariton-polariton repulsion would allow an efficient filling occurring much faster than the polariton lifetime. The elastic energy of singly quantized vortices ($l=1$) is proportional to the condensate density\cite{SmithPethick}:
\begin{equation}
E_{el}=\pi\hbar^2 n/m^\ast,
\label{Eel}
\end{equation}
and thus metallic layers acts as a potential barrier for vortices. This opens the way towards the control the VAPs trajectories by means of extra gates flanking the vortex gun. In this context, we propose first to add a pair of circular electrodes as shown in Fig.\ref{Fig3}. Owing to the symmetry of our structure, no deviation occurs until a voltage is applied to one of the two electrodes. On the other hand, if a 15 ps long Gaussian voltage pulse is imposed to one of them, the VAP deviates from its straight trajectory as it is shown at Fig.\ref{Fig3} [see Ref.\onlinecite{Movie2} for a movie]. One could of course envisage as well a configuration with a single electrode that would deviate the vortex in a deterministic way. No applied voltage would be obviously required in that case.

\begin{figure}[tbp]
\includegraphics[width=0.99\linewidth]{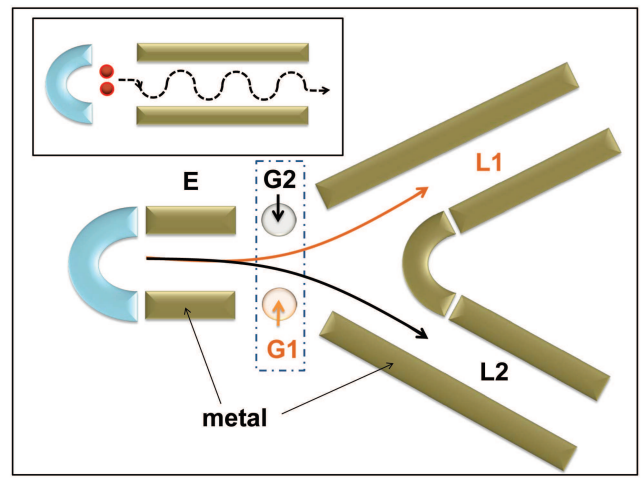}
\caption{ (Color online) \label{Fig4} Inset: Vortex guide consisting of a vortex gun and two metallic stripes. Periodic application of voltage to upper and lower stripes makes VAP to oscillate between them. Main plot: vortex switch composed of a vortex gun, a vortex guide $E$, two outgoing leads $L_1$ and $L_2$ and controlling gate electrodes $G_1$ and $G_2$. The application of the electric impulse to gates $G_1$ or $G_2$ sends VAPs to leads $L_1$ or $L_2$ respectively as marked by black and orange arrows.}
\end{figure}

Capitalizing on the previous results, we can now envisage a scheme for a vortex guide. Indeed, one can deposit two metallic stripes next to the vortex gun. The application of a sinusoidal voltage to the upper and lower stripes leads to the oscillation of the VAP between them, as it is illustrated in the inset of Fig.\ref{Fig4} [see also Ref.\onlinecite{Movie3} for a movie]. It is important that the voltage applied to the gates are weak enough not to excite parasite vortices in the system.

Vortex guides could be connected to each other and constitute a basis for varieties of logic elements. For example we can propose a design for a vortex switch shown in Fig.\ref{Fig4}: VAPs are created by the vortex gun and move along the vortex guide $E$ which splits into two outgoing leads $L_1$ and $L_2$. The metallic gates $G_1$ and $G_2$ allow a controllable redistribution of the vortices between the two paths namely: The application of the voltage impulse to the gate $G_1$ ($G_2$) sends the VAP to the outgoing lead $L_1$ ($L_2$).

\section{Vortex splitting}
\begin{figure}[tbp]
\includegraphics[width=1\linewidth]{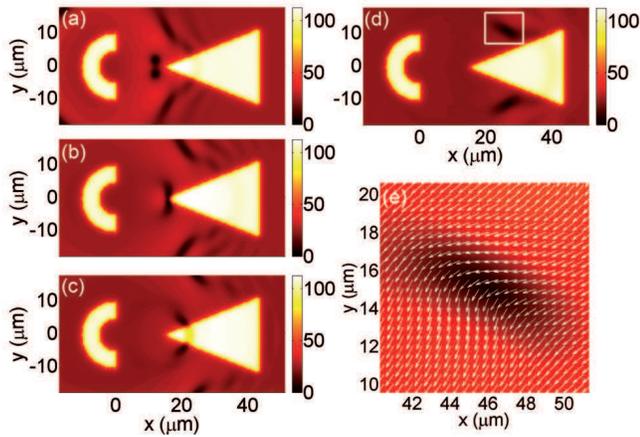}
\caption{(Color online) Scattering of a VAP induced by a metallic wedge. The panels (a)-(d) show the superfluid density at times : 10 ps, 13 ps, 15 ps and 20 ps after VAP creation respectively. The decomposition of a VAP into a pair of pseudo-topological excitations propagating along the upper and lower sides of the triangle is evidenced. The phase vector field of the upper topological excitation highlighted by the white square in (d) is shown in the panel (e). The phase of the condensate is roughly shifted by $\pi/2$ through the low density (dark) region which is characteristic for a gray soliton.}
\label{Fig5}
\end{figure}

VAPs can be unbound by means of a wedged metallic mesa, as shown in Fig.\ref{Fig5}. When a VAP reaches a tip of the wedge, it splits into two separated excitations traveling along upper and lower sides of the triangle, preserving their shape [See Ref.\onlinecite{Movie4} for a movie]. The analysis of the phase field around the upper excitations shown in the panel (e), allows to assert that they are neither individual vortices nor new VAPs. Indeed no phase singularity is visible but rather a local $\pi/2$ phase shift going through the density dip characteristic for a gray soliton or more precisely a rarefaction pulse\cite{RarefactionPulse}.

\section{Single vortices generation}
\begin{figure}[tbp]
\includegraphics[width=0.99\linewidth]{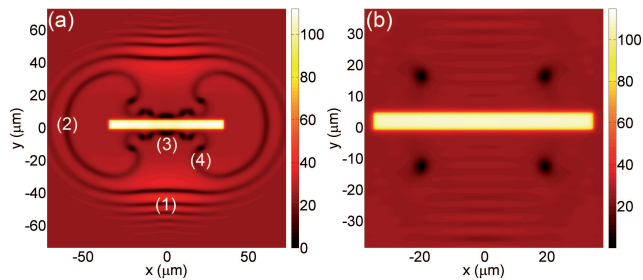}
\caption{(Color online) Generation of single vortices. (a) Superfluid density 30 ps after the arrival of the pulse which exhibit several features: (1) dispersive shock waves, (2) lasso-soliton broken into (3) VAPs and (4) single vortices. (b) Zoom on the plate region at 80 ps, the VAPs have vanished and 4 single vortices remain stable, locked at there initial positions.}
\label{Fig6}
\end{figure}

So far we have considered the nucleation of propagating vortex dipoles. But the question is: could it be possible to electrically generate single vortices in polariton condensates? To answer this question, we propose to simply use a single plate configuration as shown in the Fig.\ref{Fig6}. After the application of a short electric pulse to the plate a "lasso"-soliton is excited similarly to the one observed in Fig.\ref{Fig2}(a). It's shape is imposed by the rectangular geometry of the plate and is consequently strongly non-cylindrical. As a result the soliton is unstable and breaks into two types of excitations [see Fig.\ref{Fig6}(a)]. VAPs appear and are expelled towards the plate where they are merged, by the high density that reigns, generating phonon-like excitations. It reveals by the way an other mean of destroying a VAP, if needed, inside a circuit for example. In addition, stable single noninteracting vortices are generated at each edge of the plate as it is shown at Fig.\ref{Fig6}(b) [see Ref.\onlinecite{Movie5} for a movie]. They remain pinned at their nucleation position. In addition, dispersive shock waves induced by the local perturbation are excited in the system [see captions of Fig.\ref{Fig6}(b)].

\section{Conclusions}
We have shown that vortices can be nucleated electrically in a microcavity with metallic contacts deposited on top of the Bragg mirrors. We have proposed a structure acting as a "vortex gun" allowing on demand generation of propagating VAPs, analyzed their injection, deviation, oscillations, guiding, splitting into gray solitons. As well, we have shown the possibility to generate single vortices with a simple rectangular contact. Vortices generated in the proposed system are stable and controllable excitations that can be easily guided. Moreover, the system we propose opens the way towards the analysis of nontrivial solitonic structures as well as dispersive shock waves inside a superfluid in an unprecedent manner. The main challenge for the experimental realization of the present proposal resides in the stabilization of the system under the action of frequencies in the ranges of 10 GHz accessible nowadays\cite{Tonouchi}. We believe that it could be achieved in a near future in view of the recent technological progresses.

\section{Acknowledgements}
The work was supported by IRSES SPINMET and POLAPHEN projects and joint CNRS-RFBR PICS project. H. F. thanks International Institute of Physics, Natal, Brazil for hospitality and acknowledges the support from the FP7 ITN "Spin-Optronics" (Grant no. 237252).  M. A. K. acknowledges the support from RFBR and I. A. S. acknowledges a support from Rannis "Center of Excellence in Polaritonics".

\section{Appendix}
\begin{figure}[tbp]
\includegraphics[width=0.99\linewidth]{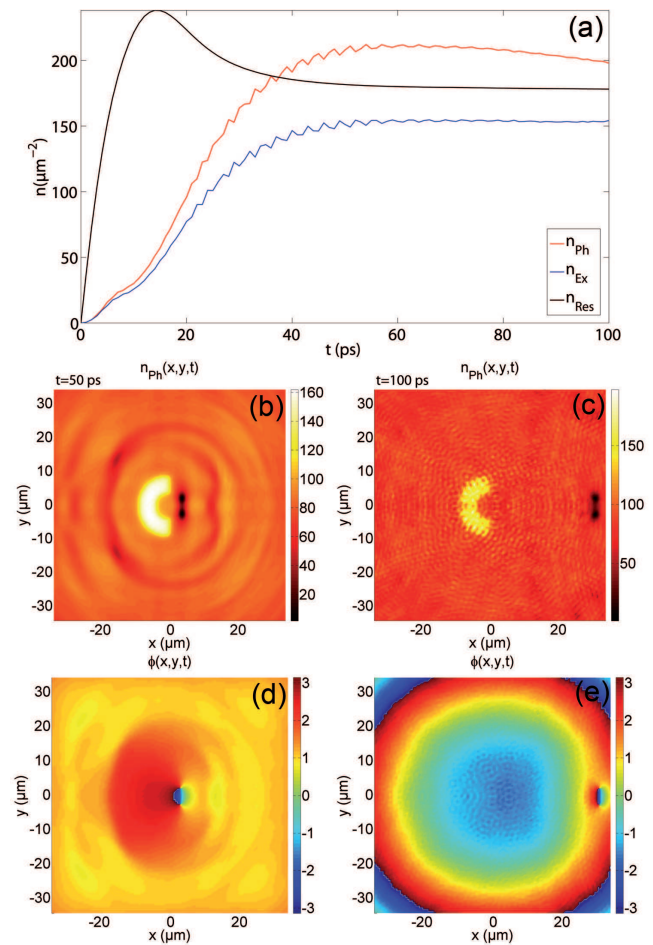}
\caption{(Color online) VAP nucleated by the application of a voltage impulse associated with a gaussian potential of amplitude $-1.5$ meV and duration 5 ps to the horseshoe contact. (a) Total densities of: the photonic component (red line), the excitonic component (blue line) and the reservoir (black line) showing the condensate being populated by the reservoir. (b) Density of the photonic component (emission out of the microcavity) 5 ps after the arrival of the impulse revealing the vortex anti-vortex pair. (c) Propagation of the VAP away from the horseshoe (gun). (d) and (e) show the phase of the photonic component associated with (b) and (c) to evidence its characteristic winding around the vortices.}
\label{FigS1}
\end{figure}
To check the validity of our model based on the equilibrium Gross-Pitaevskii equation [see eq.(2)], we have performed numerical simulations using a more accurate model: namely a set of modified Ginzburg-Landau equations. Indeed, to take into account the nonparabolicity of the polariton dispersion, we have separated the photonic field $\phi(\vec{r},t)$ from the excitonic one $\chi(\vec{r},t)$ that are coupled via the light matter interaction associated with the Rabi energy $\Omega_R=10$ meV. We have taken into account the particles lifetime: $\tau_{\phi}=15$ ps and $\tau_{\chi}=400$ ps. To model the nonresonant pumping scheme, we assumed that the exciton relaxation towards the ground state is embodied by a reservoir which evolves along a simple rate equation [eq.(1)] similarly to the proposal of Ref.\onlinecite{WCModel}. $\Gamma_R=5/\tau_\chi$ is the scattering rate between reservoir excitons and condensate polaritons. The reservoir filling is provided by the source term (pump laser) $P_R=50/\tau_R$ where $\tau_R=100$ ps is the lifetime of the particles in the reservoir. The stimulation of the condensate population is seeded by a short low amplitude probe pulse tuned at the energy of the lower polariton branch at $\vec{k}=\vec{0}$: $\omega_S=-\Omega_R$. The corresponding set of equations reads:
\begin{eqnarray}
\nonumber i\hbar \frac{{\partial \phi }}{{\partial t}} &=&  - \frac{{{\hbar ^2}}}{{2{m_\phi }}}\Delta \phi  + \frac{{{\Omega _R}}}{2}\chi\\
&+& {P_S}{e^{ - {{\left( {\frac{{t - {t_S}}}{{\Delta {t_S}}}} \right)}^2}}}{e^{ - i{\omega _S}t}}
- \frac{{i\hbar }}{{2{\tau _\phi }}}\phi \\
\nonumber i\hbar \frac{{\partial \chi }}{{\partial t}} &=&  - \frac{{{\hbar ^2}}}{{2{m_\chi }}}\Delta \chi  + \frac{{{\Omega _R}}}{2}\phi  + \alpha \left( {{{\left| \chi  \right|}^2} + {n_R}} \right)\chi \\
&+& \left( {{U_0} + {U_V}} \right)\chi - \frac{{i\hbar }}{{2{\tau _\chi }}}\chi  + i\hbar \frac{{{\Gamma _R}}}{2}{n_R}\chi \\
\frac{{\partial {n_R}}}{{\partial t}} &=& {P_R} - \frac{{{n_R}}}{{{\tau_R}}} - {\Gamma _R}{\left| \chi  \right|^2}{n_R}
\end{eqnarray}

In this framework, we present the result of a numerical simulation in the figure \ref{FigS1} showing the vortex gun configuration (see captions). As one can see, the results obtained remain very similar to the one obtained in the main text using the simplified approach [to be compared with the Fig.\ref{Fig2}]. The main difference is that the results are less smooth and therefore less instructive due to some instabilities, intrinsic to the model, that develop at later times [visible in the density landscape of the panel (e)]. We therefore assert that the simple model used in the main text is sufficient to realistically describe the vortex nucleation and guiding processes.

\end{document}